%% file: main.tex
\documentclass[conference]{IEEEtran}
\IEEEoverridecommandlockouts

\usepackage{cite}
\usepackage{amsmath,amssymb,amsfonts}
\usepackage{algorithmic}
\usepackage{graphicx}
\usepackage{textcomp}
\usepackage{comment}
\usepackage{xcolor}
\usepackage{tabularx}
\usepackage{caption}
\usepackage{subcaption}
\usepackage{subcaption}
\usepackage{array, makecell}
\usepackage[bookmarks=false]{hyperref}
\usepackage{url}
\usepackage{soul}

\setcellgapes{2pt}

\usepackage{pifont}
\newcommand{\cmark}{\ding{51}}%
\newcommand{\xmark}{\ding{55}}%

\def\BibTeX{{\rm B\kern-.05em{\sc i\kern-.025em b}\kern-.08em
    T\kern-.1667em\lower.7ex\hbox{E}\kern-.125emX}}

\begin{document}

\title{tf-Darshan: Understanding Fine-grained I/O Performance in Machine Learning Workloads}

\author{\IEEEauthorblockN{Steven W. D. Chien\IEEEauthorrefmark{1}, Artur Podobas\IEEEauthorrefmark{1}, Ivy B. Peng\IEEEauthorrefmark{2}, Stefano Markidis\IEEEauthorrefmark{1}} 
	\IEEEauthorblockA{
		\textit{\IEEEauthorrefmark{1} Division of Computational Science and Technology}, KTH Royal Institute of Technology, Stockholm, Sweden \\
		\textit{\IEEEauthorrefmark{2} Lawrence Livermore National Laboratory}, Livermore, USA \\
		\IEEEauthorrefmark{1}\{wdchien,podobas,markidis\}@kth.se, \IEEEauthorrefmark{2}peng8@llnl.gov
	}
}

\maketitle

\begin{abstract}
Machine Learning applications on HPC systems have been gaining popularity in recent years. The upcoming large scale systems will offer tremendous parallelism for training through GPUs. However, another heavy aspect of Machine Learning is I/O, and this can potentially be a performance bottleneck. TensorFlow, one of the most popular Deep-Learning platforms, now offers a new profiler interface and allows instrumentation of TensorFlow operations. However, the current profiler only enables analysis at the TensorFlow platform level and does not provide system-level information. In this paper, we extend TensorFlow Profiler and introduce tf-Darshan, both a profiler and tracer, that performs instrumentation through Darshan. We use the same Darshan shared instrumentation library and implement a runtime attachment without using a system preload. We can extract Darshan profiling data structures during TensorFlow execution to enable analysis through the TensorFlow profiler. We visualize the performance results through TensorBoard, the web-based TensorFlow visualization tool. At the same time, we do not alter Darshan's existing implementation. We illustrate tf-Darshan by performing two case studies on ImageNet image and Malware classification. We show that by guiding optimization using data from tf-Darshan, we increase POSIX I/O bandwidth by up to 19\% by selecting data for staging on fast tier storage. We also show that Darshan has the potential of being used as a runtime library for profiling and providing information for future optimization.
\end{abstract}


\input{introduction}
\input{background}
\input{methods}
\input{evaluation}
\input{results}
\input{rel-work}
\input{conclusion}

\section*{Acknowledgment}
\scriptsize{Funding for the work is received from the European Commission H2020 program, Grant Agreement No. 800999 (SAGE2). Experiments were performed on resources provided by the Swedish National Infrastructure for Computing (SNIC) at HPC2N. LLNL-CONF-810737.}

\bibliographystyle{IEEEtran}
\bibliography{main}
\end{document}

%% file: introduction.tex
\section{Introduction}
\label{sec:introduction}

Machine Learning (ML) has gained considerable attention in recent years for its potential to drive advances in self-driving cars, drug development, and image processing. Demands for ML workloads on many HPC systems are increasing overtime~\cite{patel2019revisiting,10.1145/3295500.3356157}. New ML methods, such as Deep-Learning (DL)~\cite{lecun2015deep}, involve a computational intensive minimization process. Special devices, such as Google TPU~\cite{tpu} and NVIDIA TensorCore~\cite{markidis2018nvidia}, are designed to boost the computation performance of DL workloads. Nevertheless, ML workloads are also data-intensive. Massive datasets are required to train an accurate ML model. The high-pace data ingestion heavily stresses the I/O system. Previous works~\cite{10.1145/3331526, chien2018characterizing, zhumulti, 8526881} have characterized I/O performance in large-scale ML workloads and showed that without an efficient data preprocessing pipeline, ML workloads are highly input-bound. 

Pinpointing I/O bottlenecks in ML workloads is crucial for enabling efficient data preprocessing. Currently, popular DL frameworks (such as TensorFlow) lacks a much-needed tool that would enable application developers to understand fine-grained I/O performance. Application developers are responsible for obtaining I/O characterization to select appropriate optimization techniques. Although the latest release of TensorFlow 2.2.0 provides a new TensorFlow Profiler~\cite{tfprofiler}, the profiler can only provide coarse-grained information on compute operations. For instance, the TensorFlow Profiler can profile and trace computing operations on the TensorFlow platform to provide a high-level view of the training process, such as time for reading a file. However, less focus is on fine-grained I/O details. Fine-grained system-level information could provide a complete and detailed picture, allow a better understanding of performance observations, and guide the optimization of I/O operations. In this work, we provide a tool called tf-Darshan to extend the TensorFlow Profiler for fine-grained I/O characterization. 

tf-Darshan, a metric rich TensorFlow profiler, uses the capability of Darshan~\cite{carns200924, 10.1145/2027066.2027068, xu2017dxt, 8752753, patel2020uncovering},  a popular I/O profiler, to enable profiling and tracing of I/O operations in ML workloads in the TensorFlow framework. tf-Darshan provides runtime attachment of I/O instrumentation. It also supports profiling and trace data management. Results from tf-Darshan can be directly visualized through TensorBoard.

The main contributions of this paper are as follows:
\begin{enumerate}
\item We introduce the latest TensorFlow Profiler and illustrate its feasibility for supporting new profilers within TensorFlow.
\item We design and implement \textit{tf-Darshan} for fine-grained I/O characterization in DL workloads.
\item We identified and adapted the key data structures in Darshan for enabling data extraction and tracing during the runtime of ML workloads
\item We quantify the overhead and validate the correctness of tf-Darshan
\item We demonstrate optimizations guided by information from tf-Darshan in two case studies on ImageNet Classification and Malware Detection, using an HPC system and a workstation with fast storage respectively.
\end{enumerate}

%% file: background.tex
\section{Background and Motivation}
\label{sec:background}
Improvement towards ML computation workload has been well studied in recent years. For example, the NVIDIA Volta GPU architecture, features TensorCore, a specialized core that performs GEMM operation in mixed-precision, targeting one of the most computationally intensive parts of the workload. Apart from being computationally intensive, ML is also data-intensive, requiring a large number of data samples for training purposes. In other words, it becomes an input-bound application if the I/O system does not keep up with the high computational performance. Previous studies have shown that I/O can account for as much as 90\% of the total training time~\cite{10.1145/3331526}. Unlike traditional HPC collective I/O, where processes rearrange I/O operations through a communicator to maximize bandwidth and minimize metadata operation, ML I/O uses an independent I/O strategy. Distributed training in ML is often implemented using data-parallelism, where workers individually train a network with different samples. During training, their weights are synchronized periodically. ML frameworks attempt to solve the I/O issue in different ways. For example, TensorFlow~\cite{abadi2016tensorflow,chien2019tensorflow}, a leading ML framework, introduces the \textsf{\small{tf.data}} API~\cite{tf.data}, for implementing I/O and preprocessing pipelines, allowing optimizations such as parallel execution of the pipelines and prefetching of data to overlap data preprocessing on CPU and training on GPU~\cite{8526881,chien2018characterizing}. \textsf{\small{tf.data}} enables parallel input pipeline execution using \textsf{\small{tf.data.map}} with a capture function enabling input I/O and preprocessing on different threads. A user can specify the number of pipelines to execute concurrently with the parameter \textsf{\small{num\_parallel\_calls}}, either setting a number or using \textsf{\small{tf.data.experimental.AUTOTUNE}} to let the TensorFlow's optimizer determine the number of threads automatically. When a pipeline is being invoked, a thread pool will be used to execute the capture functions in parallel. To exploit concurrency between the CPU and GPUs, \textsf{\small{tf.data}} allows prefetching, to continue executing the input pipeline while training is still in-progress on the GPU. Data batches will be accumulated in a buffer that will be readily available for the next step of training immediately when the GPU completes the current batch.

To understand and characterize the underlying I/O fine-grained operations is essential to optimize an ML pipeline. While GPU and CPU profilers are widely available, non-trivial integration with the specific ML platform is required to infer specific workload performance. For this reason, the recent release of TensorFlow 2.2.0 introduces a new TensorFlow Profiler that provides detailed information on compute workload performed on the TensorFlow platform. Furthermore, the TensorBoard web interface allows for visualizing the performance results. TensorFlow Profiler implements host-side tracing through the execution engine, and GPU tracing through CUDA Profiling Tools Interface (CUPTI)~\cite{cupti}. For example, we can extract the performance of individual computation kernels and detailed tracing information. This information collected during execution can be useful for performing optimization in the future.

In terms of data preparation, TensorFlow Profiler provides an Input-pipeline analysis that quantifies steps taken in the \textsf{\small{tf.data}} pipeline. However, the information provided remains on a TensorFlow level, meaning that it does not provide fine-grained system-level information, such as POSIX operations. Information such as POSIX read size on files and time spent on metadata operation can be important for optimizations such as data placement on multi-tiered I/O systems. We aim to fill this gap by implementing tf-Darshan to provide fine-grained profiling and tracing of the I/O workload. tf-Darshan relies on Darshan, one of the most comprehensive and used HPC I/O workload profiler and tracer. Darshan is a transparent low-overhead profiler that is implemented as a shared library and defers statistical post-processing operations after the application returns. Because Darshan completes the analysis in the post-processing step, to extract information of application I/O operations and perform an analysis at runtime is not possible. For this reason, we implement tf-Darshan with a loosely coupled integration with the Darshan shared library through a runtime attachment. We also augment Darshan to allow the returning of internal data structures to the instrumented application. In other words, the instrumented application can inspect statistics collected by Darshan and perform an analysis at runtime. This capability could provide an opportunity to perform runtime optimization, based on Darshan's data.

\begin{table}[t]
	\centering
	\caption{Comparison of Darshan and tf-Darshan for profiling TensorFlow workloads.\label{tab:darshan-vs-tfdarshan}}
	\begin{tabular}{|l|c|c|}
		\hline
		\textbf{Feature} & \multicolumn{1}{l|}{\textbf{Darshan}} & \multicolumn{1}{l|}{\textbf{tf-Darshan}} \\ \hline
		Modules & POSIX, STDIO, DXT & POSIX, STDIO, DXT \\ \hline
		Transparent & \cmark & \cmark \\ \hline
		Runtime start/stop & \xmark & \cmark \\ \hline
		Log analysis & Post-execution & In-situ \\ \hline
		Reporting & \begin{tabular}[c]{@{}c@{}}After whole\\ application returns\end{tabular} & After Profiling stops \\ \hline
		Outputs & Darshan log & Darshan log, Protobuf \\ \hline
		Visualization & PDF, Log utilities & TensorBoard web \\ \hline
	\end{tabular}
\end{table}

While tf-Darshan uses the same Darshan logging capabilities, there are several important operational differences. We summarize the differences in Table.~\ref{tab:darshan-vs-tfdarshan}.

%% file: methods.tex
\section{Methodology}
\label{sec:methods}

In this section, we introduce and describe tf-Darshan-- a tool for profiling I/O performance of ML-applications in state-of-the-art HPC and data-centers. tf-Darshan consists of three distinct components:
\begin{itemize}
\item a specialized tracer that for use within the TensorFlow runtime system, 
\item a customized wrapper that attaches and interfaces Darshan instrumentation/data extraction functionality from within the TensorFlow runtime system and fully transparently to the user, and
\item a plugin that allows users to intuitively visualize tf-Darshan I/O profiles and results using the TensorBoard interface.
\end{itemize}

Our contribution, tf-Darshan, is based on a new version of Darshan that allows decoupling with MPI interfaces and pure POSIX (non-MPI) instrumentation. We based our design on this version (3.2.0-pre~\cite{darshan-serial}) since TensorFlow is not an MPI application. Since we adopt a loose integration of Darshan instead of a complete embedding of Darshan into TensorFlow, it is possible to use newer versions of Darshan as long as the augmented data extraction functions are provided. The parallel version of Darshan uses the PMPI profiling interface to intercept MPI-IO calls in applications. If TensorFlow employs MPI as a distributed strategy for I/O in the future, one can employ the parallel version of Darshan with the MPI module to profile and instrumentation I/O activities with a similar technique.

\subsection{TensorFlow Profiler}

Our contribution, tf-Darshan, is based on and extends the functionality of the recently added interface in TensorFlow 2.2.0, which we will briefly describe herein.

\begin{figure}[t]
	\centering
	\includegraphics[width=\columnwidth]{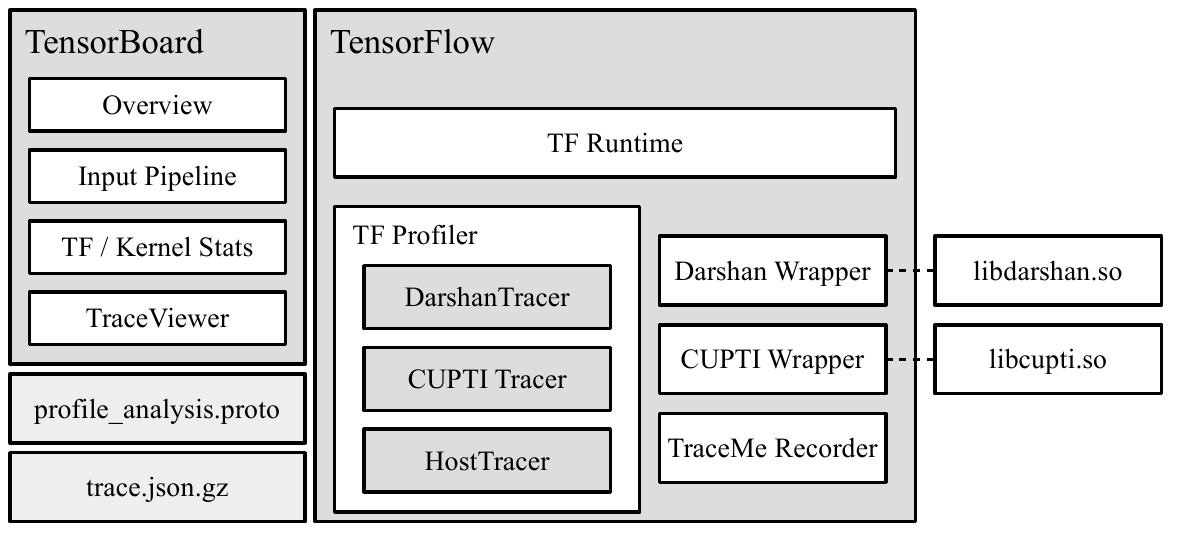}
	\caption{Organization of TensorFlow Profiler and tools.}
	\label{fig:tf-profile-overview}
\end{figure}

The recent TensorFlow 2.2.0 implementation provides two key features for profiling and tracing ML workloads: \textbf{(i)} functionality for collecting traces and runtime systems from the host CPUs and discrete GPUs, and \textbf{(ii)} visualization of gathered information intuitively through on an interactive dashboard. Fig.\ref{fig:tf-profile-overview} provides an overview of the organization of the tools.

In Tensorflow 2.2.0, a component called a \textit{tracer} performs the collection and gathering of runtime information. A tracer provides a shared API for data acquisition and collection to the Tensorflow runtime system, which manages and invokes the tracer at a specific point during execution (depending on what the runtime system wants to profile). While the tracer does provide a unified API for managing data, it outsources the data collection (which can be very hardware-specific) to other components. For example, for profiling GPU-related information, the TensorFlow 2.2.0 tracer would rely on the CUPTI~\cite{cupti} library to provide low-level GPU traces and characteristics. For host-related (CPU) profiling, the tracer uses another module (a background recorder) to capture CPU information. The TensorFlow tracers later relay the information obtained by the hardware-specific profilers to the TensorFlow runtime system. Finally, the TensorFlow runtime system will format, prepare, and output the data through protocol buffers (protobuf) for later visualization.

Users currently have three options for invoking and using TensorFlow 2.2.0 tracers:
\begin{itemize}
\item \textbf{Automatically}: users can automatically enable tracers from existing APIs (such as Keras) by providing a TensorBoard callback hook, which then operates on a batch-level. Currently, it supports the specification of one range of batches,
\item \textbf{Manually}: users can manually decide what to profile by inserting calls to the TensorFlow profiler (\textsf{\small{tf.profiler.experimental.start()/stop()}}), allowing more fine-grained control over what is being profiled, and
\item \textbf{Interactively}: users can interactively start and stop tracing from a different machine by allowing TensorBoard to connect to TensorFlow through a network socket (\textsf{\small{tf.profiler.server.start()}})
\end{itemize}

In tf-Darshan, we support all three options for using tracers and we leverage TensorFlow 2.2.0's modular and implementation-agnostic tracer design to extend the limited functionality that it currently provides. These capabilities are facilitated by the TensorFlow runtime system, as long as we provide a new interface for starting/stopping the profiler and collecting the data.

\subsection{Runtime attachment of instrumented functions}

\begin{figure}[t]
    \centering
    \includegraphics[width=\columnwidth]{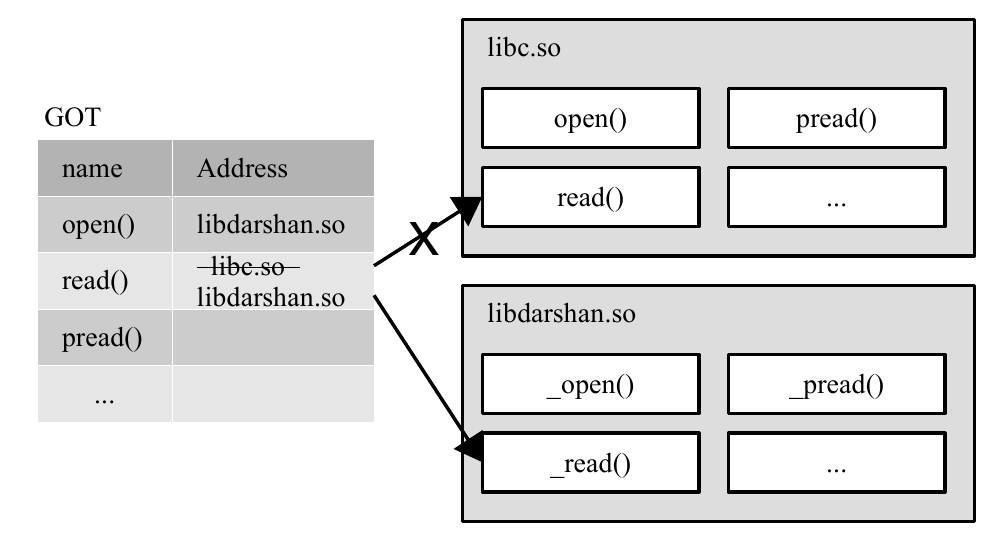}
    \caption{Performing runtime replacement of function addresses by patching the Global Offset Table.}
    \label{fig:tf-darshan-runtime-instrumentation}
\end{figure}

Our aspiration with tf-Darshan is to provide users with intuitive and interactive visualization (through TensorBoard) that leverage existing and mature HPC tracing software (Darshan).  Today, Darshan is implemented as a shared library. Profiling using Darshan is performed by overloading (e.g., through the \texttt{LD\_PRELOAD} environment variable) the different I/O functions (e.g., \texttt{fread} or \texttt{fwrite}) that usually go to a standard library (e.g., \emph{libc}) to instead go through Darshan. This approach is easy to use and also transparent to the user, but it has one drawback: the profiled application cannot easily access the profiled data, which is a prerequisite to (for example) auto-tune parameters (e.g., prefetch or migrating files). Instead, tf-Darshan provides a \textit{middle man} to facilitate the free transfer of profiled information across both the profiler and the profiled application. 

The tf-Darshan middle-man component is responsible for the communication between the TensorFlow layer and then Darshan layer. When a TensorFlow tracer starts, our middle-man dynamically (through \texttt{dlopen}) loads the Darshan shared library into TensorFlow's memory. Next, we scan the Global Offset Table (GOT) in search of any symbols that we are interested in profiling. These symbols (e.g., \texttt{fread} or \texttt{fwrite}) usually point to standard external libraries (such as \emph{libc} or \emph{newlibc}). Upon finding candidates for profiling, the middle-man patches the table and redirects these symbols to the Darshan shared library instead, granting Darshan full control over the I/O access. This mechanism is illustrated in Fig.~\ref{fig:tf-darshan-runtime-instrumentation}. From this point in the application, all subsequent I/O operations pass through the Darshan profiler. This technique of dynamically patching the GOT is in use in other contexts, such as for runtime debugging (e.g. Heaptrack~\cite{Kde2020May}), and allow bi-directional information flow between the TensorFlow and Darshan systems.

\subsection{Extraction of Profiling Results}
To extract information from Darshan, we implemented several data extraction functions in the Darshan shared library that returns Darshan module buffers that record counters for I/O operations to individual files. We also export other helper functions such as file name lookup (through \texttt{dlsym}). tf-Darshan's wrapper does not only handle all the symbol management but also on profile data management. When a profiling session begins, our tracer calls the wrapper to make a copy of the Darshan module data structures. tf-Darshan makes another copy of the module data structures when a profiling session stops. When the TensorFlow runtime collects profiling data, the two samples collected during start and stop are analyzed by tf-Darshan to retrieve relevant statistics. For each file recorded by Darshan, we extract the detailed tracing information and export them as trace events.

\subsection{TensorBoard Integration}
The information collected by the TensorFlow runtime will be converted and exported as Protobuf files for visualization. We modified the TensorBoard Profile plugin in order to visualize additional results provided by Darshan. We expand the Input-Pipeline Analysis section to include various statistics. For instance, POSIX bandwidth, the number of POSIX I/O operations performed, distribution of POSIX read size, and distribution of file size. Furthermore, individual POSIX read and write operations on different files can be visualized as a timeline by the TraceViewer tool.

%% file: evaluation.tex
\begin{table*}[t]
    \centering
    \caption{Characteristics of datasets and configurations used in the test cases.\label{tab:use-case-summary}}
    \makegapedcells
    \resizebox{\textwidth}{!}{
        \begin{tabular}{|l|l|l|l|l|l|l|l|l|l|l|}
            \hline
            {\color[HTML]{000000} \textbf{Name}} &
            {\color[HTML]{000000} \textbf{Description}} &
            {\color[HTML]{000000} \textbf{Batch size}} &
            {\color[HTML]{000000} \textbf{Steps}} &
            {\color[HTML]{000000} \textbf{Threads}} &
            {\color[HTML]{000000} \textbf{Prefetch}} &
            {\color[HTML]{000000} \textbf{No. Files}} &
            {\color[HTML]{000000} \textbf{Total Size}} &
            {\color[HTML]{000000} \textbf{Median Size}} &
            {\color[HTML]{000000}\textbf{System}} &
            {\color[HTML]{000000} \textbf{Character}} \\ \hline\hline
            {\color[HTML]{000000} STREAM (ImageNet,Malware)} &
            {\color[HTML]{000000} Fetching data without compute} &
            128 &
            100,50 &
            16 &
            10 &
            {\color[HTML]{000000} 12,800, 6400} &
            {\color[HTML]{000000} $\sim$1~GB, $\sim$35~GB} &
            $\sim$76~KB, $\sim$7.3~MB &
            Greendog &
            No preprocessing to validate bandwidth. \\ \hline
            {\color[HTML]{000000} Kaggle BIG 2015 Challenge} &
            {\color[HTML]{000000} Malware classification} &
            32 &
            339 &
            1, 16 &
            10 &
            10868 &
            {\color[HTML]{000000} $\sim$48~GB} &
            $\sim$4 MB &
            Greendog &
            Large individual files. \\ \hline
            {\color[HTML]{000000} ImageNet } &
            {\color[HTML]{000000} Classifying images } &
            256 &
            500 &
            1, 2 &
            10 &
            {\color[HTML]{000000} 128,000 } &
            {\color[HTML]{000000} $\sim$11.6~GB } &
            $\sim$88 KB &
            Kebnekaise &
            Large number of small files. \\ \hline
        \end{tabular}%
    }
\end{table*}

\section{Evaluation}
\label{sec:evaluation}
In this section, we introduce the experimental setup and evaluated applications. We demonstrate three key capabilities of tf-Darshan -- in-situ profiling during training sessions, visualization at TensorBoard, and low overhead.

\subsection{Experimental Setup}

We evaluate the features of tf-Darshan on two platforms:

\noindent - \textbf{Greendog} is a workstation with an i7-7820X processor (8 cores), 32~GB DRAM, and one NVIDIA RTX2060 SUPER GPU. The workstation features three types of storage: two 2~TB HDD (non-RAID), one 1~TB SSD, and one 480~GB Intel Optane SSD 900p on PCIe. The file system used is ext4, and the OS is Ubuntu 18.04 LTS. TensorFlow is compiled with GCC v7.5.0 and NVIDIA CUDA 10.2. We store all our datasets on the HDD.

\noindent - \textbf{Kebnekaise} is an HPC cluster at HPC2N in Umeå. We use a computing node with two Intel Xeon Gold 6132 CPUs (28 cores), 192~GB RAM, and two NVIDIA V100 GPUs on PCIe. The cluster is connected with EDR Infiniband and uses a Lustre parallel file system. The OS is Ubuntu 16.04.6 LTS. TensorFlow is compiled with GCC v8.3.0, NVIDIA CUDA 10.1, and NVIDIA NCCL 2.4.8.

For each experiment on \textit{Greendog} where we have root permission, we drop the page cache before every benchmark to eliminate the interference from cached pages in memory. Only one epoch per experiment is performed to avoid reusing cached data in the second epoch. We fix the CPU frequency to the highest by running \texttt{cpupower frequency-set --governor performance} to avoid performance variability~\cite{10.1145/3297663.3310299}.

We first validate the profiling results from tf-Darshan using a STREAM benchmark and comparing them to the background I/O obtained from dstat~\cite{dstat-real2019Jan}. Then, we quantify the overhead of tf-Darshan when compared with a training session without profiling. Finally, we perform two case studies using tf-Darshan. They include two ML applications that are implemented in Keras, which both use the SGD optimizer with default parameters (learning rate=0.01, momentum=0.0) and categorical cross-entropy as the loss function. Their details are as follow:

\noindent - \textbf{ImageNet Classification} classifies images in ImageNet~\cite{imagenet11} by training an AlexNet with a batch size of 256 and we run 500 steps, without data augmentation. We use \textsf{\small{tf.data}} API to implement the data pipeline with \textsf{\small{tf.data.map}}. Within the capture function, I/O is performed using \textsf{\small{tf.io.read\_file}}, and then preprocessing (decode, resize, batching) is performed.

\noindent - \textbf{Microsoft Malware Detection Challenge 2015} classifies malware Byte code by training a simple two-layer Convolution Neural Network. The dataset is a part of the Kaggle BIG 2015 Challenge~\cite{ronen2018microsoft} and consists of nine malware classes with approximately 10,000 Byte code samples. Similar to the ImageNet Classification, we implement the data pipeline using \textsf{\small{tf.data}} API. We read the byte code files and decode them as images before feeding them into the neural network for training.

We summarize their details as the following and the configurations in Table \ref{tab:use-case-summary}.

\subsection{Tool Validation}

\begin{figure}[t]
    \centering
    \includegraphics[width=\linewidth]{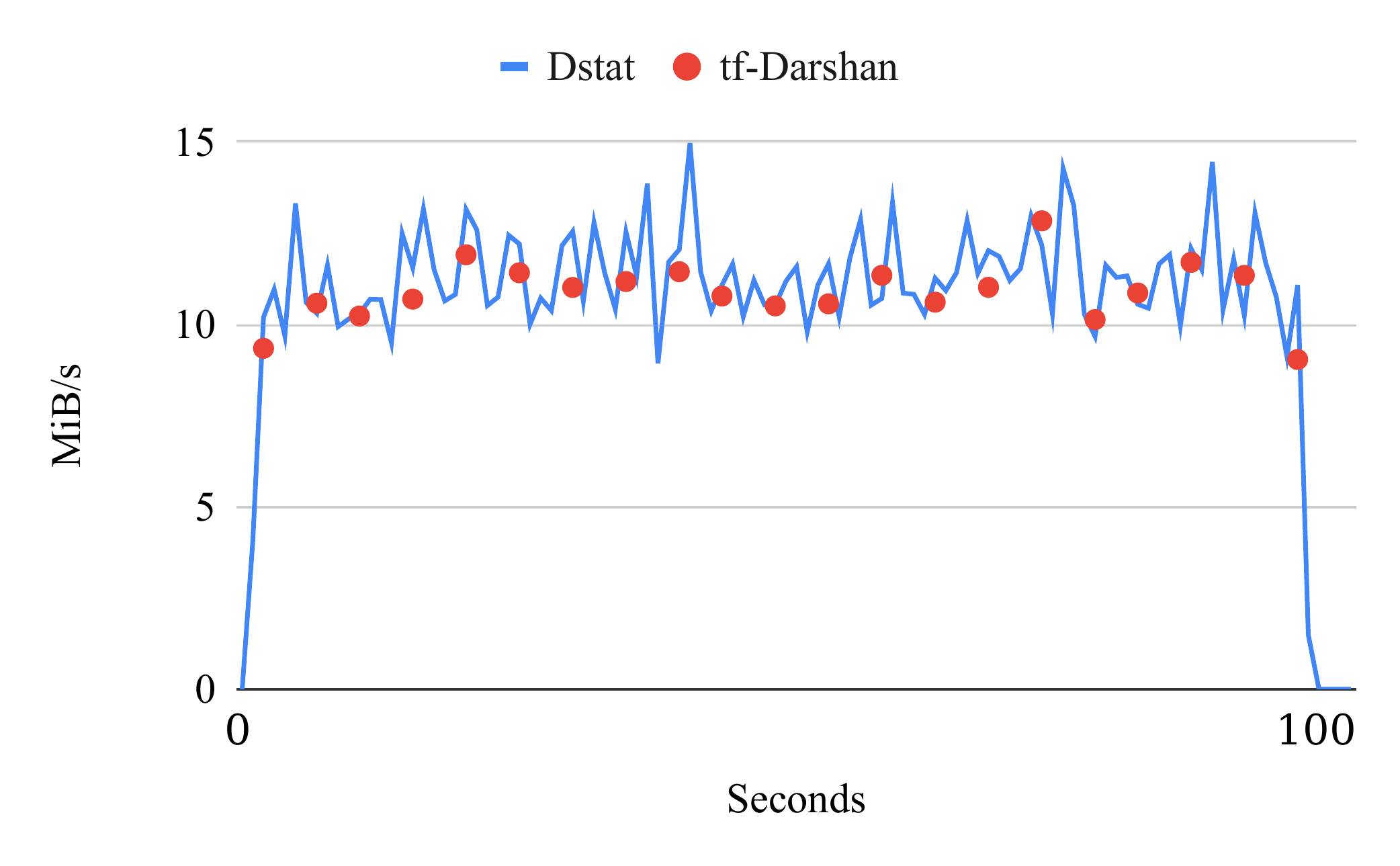}
    \caption{STREAM(ImageNet) bandwidth measured using batch size 128 for 100 steps.}
    \label{fig:imagenet-bandwidth}
\end{figure}

\begin{figure}[t]
    \centering
    \includegraphics[width=\linewidth]{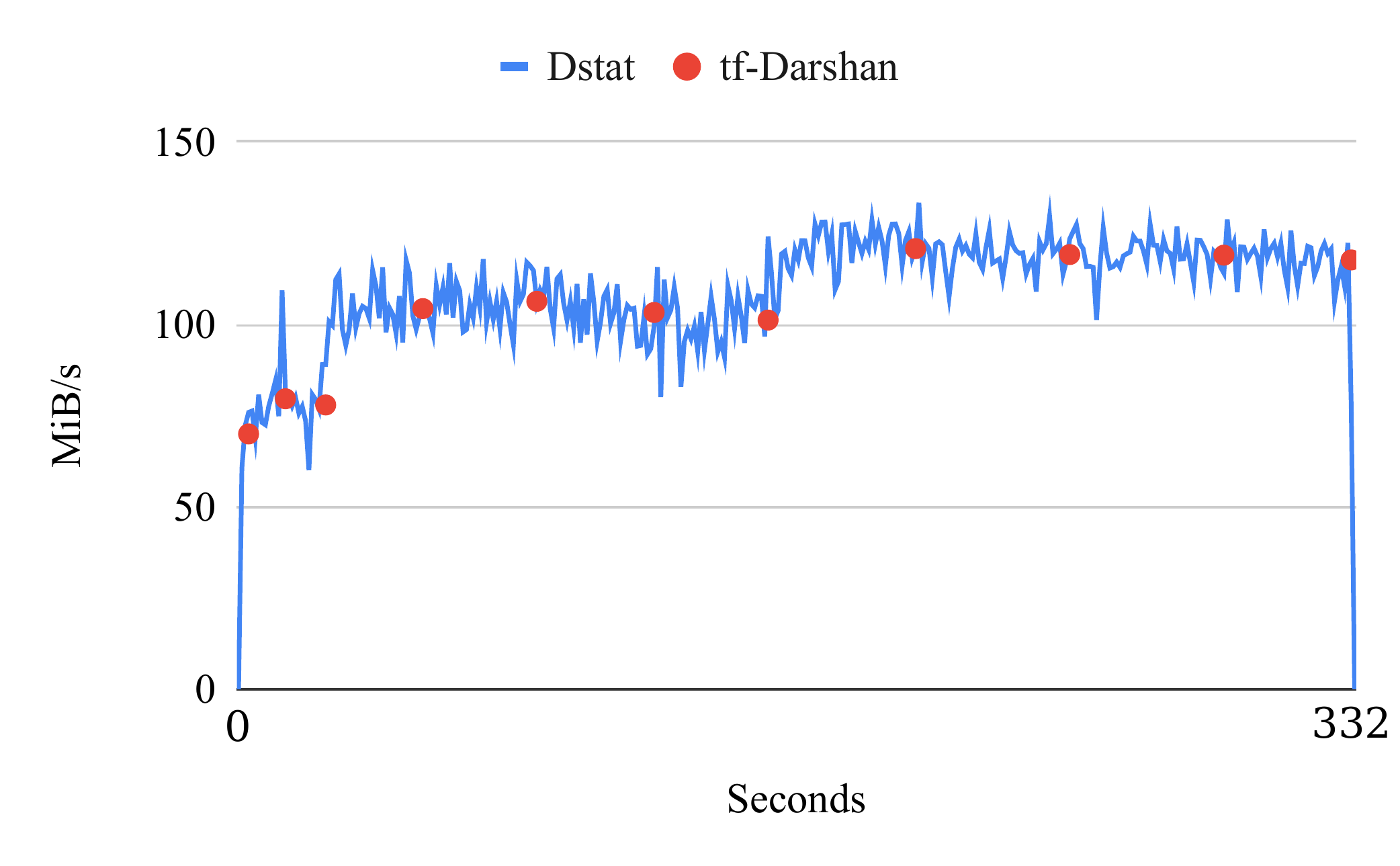}
    \caption{STREAM(Malware) bandwidth using batch size 128 for 50 steps.}
    \label{fig:malware-bandwidth}
\end{figure}

Bandwidth is one of the most critical metrics in I/O performance characterization. We compute the bandwidth of training in tf-Darshan by the total number of bytes transferred and elapsed wall-clock time during the profiling session. We compare the total I/O sampled at the beginning and at the end of profiling to derive the amount of I/O during the profiling session. We examine the raw I/O capability of TensorFlow's \textsf{\small{tf.data}} API by implementing a STREAM application that performs no computation and preprocessing other than reading files and forming batches. We run the STREAM benchmark using the datasets from the ImageNet and Malware classification and present the results in MB/s. Both cases use a batch size of 128, 16 I/O threads, and prefetching of 10 batches. We run the benchmark for 100 steps with ImageNet case and 50 steps with Malware use-case. Furthermore, we stop and restart profiling in TensorFlow every five steps to derive a bandwidth. To validate the bandwidth value derived by tf-Darshan, we concurrently run Dstat in the background to collect disk activities.

We present the benchmark results in Fig.~\ref{fig:imagenet-bandwidth} for ImageNet and Fig.~\ref{fig:malware-bandwidth} for Malware use-case respectively. The blue lines represent the runtime bandwidth measured by Dstat each second, while the red dots represent the bandwidth measured tf-Darshan after every five batches. We note that tf-Darshan determines the I/O bandwidth at high-accuracy. Another observation is that the bandwidth in our malware use-case is approximately $10\times$ higher than in ImageNet, which can be explained by small file sizes in the ImageNet dataset.

\subsection{Overhead}

\begin{figure}[t]
    \centering
    \includegraphics[width=\linewidth]{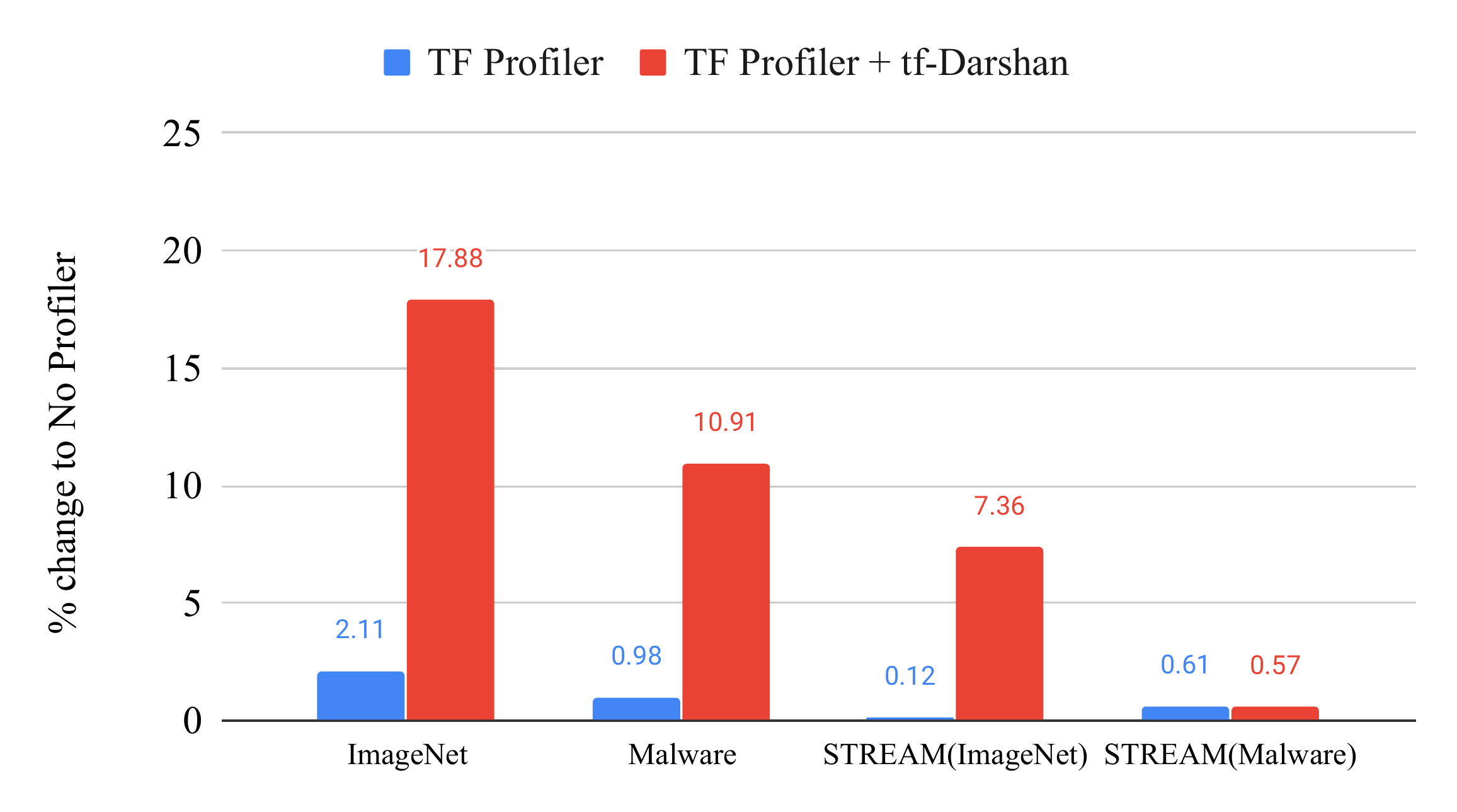}
    \caption{Training and streaming performance when the automatic TensorBoard callback and manual trigger of profiling is used. Error bar is omitted as standard deviation is below $2\%$.}
    \label{fig:overhead}
\end{figure}

A common issue with fine-grained profiling and tracing is the overhead to application execution. Since our instrumentation is based on Darshan, the overhead introduced during application execution is negligible. We quantify the overhead by running our two use-cases five times with a batch size of 128 and 10 steps, as well as the Stream benchmarks, and report the average execution time. We perform profiling over the entire 10 steps, meaning that we start from the first batch and stop at the last. By measuring the time needed for Keras' model fitting to return, we show that our tracer introduces overhead approximately between 10\% to 20\% compared to when no profiler is used (Fig.~\ref{fig:overhead}). However, much of the overhead occurs during trace data collection and analyzes after profiling stops. This can be verified by the background disk activity (Fig.~\ref{fig:malware-dstat}) from Section~\ref{sec:case-studies}, that tf-Darshan introduces an overhead at the end of the training when it collects all the traces and statistics. While our use-cases use TensorBoard callback in Keras to profile automatically, we repeat the Stream benchmark using the manual profiling method and restart profiling every five steps. We get approximately 0.6\% to 7\% overhead, relative to when no profiling is used. One observation is that the overhead has a strong correlation against the number of files processed. Apart from trace data analysis after profiling ends, a possible reason is the extra operations introduced during tracing. Internally, Darshan stores a set of counters for each and every file that the application interacts with. In other words, the number of operations increases by the number of files processed in combination with the I/O operations performed. Furthermore, tf-Darshan performs in-situ log analysis after profiling stops, of which a large number of files may lead to higher overhead.
\subsection{Checkpoint}

\begin{figure}[t]
	\centering
	\includegraphics[width=\linewidth]{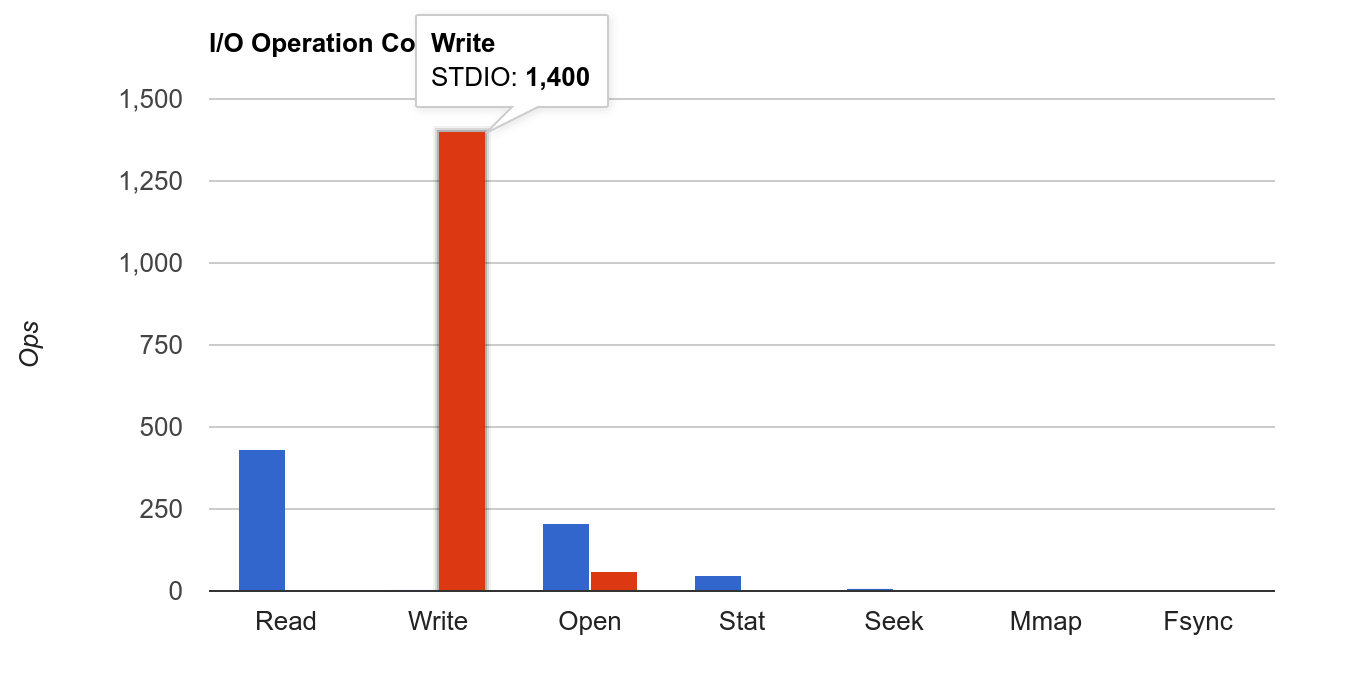}
	\caption{tf-Darshan capturing write activities on STDIO layer.}
	\label{fig:checkpoint}
\end{figure}

Checkpointing in TensorFlow is another functionality that uses I/O extensively, besides data ingestion. Checkpoints in ML workloads serve two primary purposes, i.e., saving a state for restart after a failure, and examining the internal state of a network during training. In TensorFlow, a checkpoint captures the values of all the variables inside a model. To recover from a checkpoint, values in the checkpoint are loaded into the original model. Checkpoints are typically captured periodically, either by steps or by epochs. In Keras, this can be specified using the \textsf{\small{tf.keras.callbacks.ModelCheckpoint}} callback; in case of custom training loops, \textsf{\small{tf.train.Checkpoint}} is used to capture a checkpoint, and multiple checkpoints can be managed by \textsf{\small{tf.train.CheckpointManager}}. In the TensorFlow POSIX file system module, writable files are written through \texttt{fwrite} in STDIO. Therefore, Darshan's STDIO module can also capture these write activities if required.

We illustrate this with a small example. We train our image classification use-case with  10 steps, making a checkpoint made after every step. We keep all the checkpoints so 10 checkpoints will be created. As our ML applications are implemented in Keras, we use the \textsf{\small{tf.keras.callbacks.ModelCheckpoint}} callback to specify the configurations. As TensorFlow uses \texttt{fwrite} when writing data, the result is shown in STDIO layer as seen in Fig.~\ref{fig:checkpoint}, where 1,400 calls to \texttt{fwrite} have been made.

%% file: results.tex
\section{Case Studies}
\label{sec:case-studies}

In this section, we characterize the I/O performance of the selected use-cases using tf-Darshan and show how the information from tf-Darshan can guide optimization.

\subsection{Image Classification}
\begin{figure*}[t]
	\centering
	\begin{subfigure}{\linewidth}
		\centering
		\includegraphics[width=\linewidth]{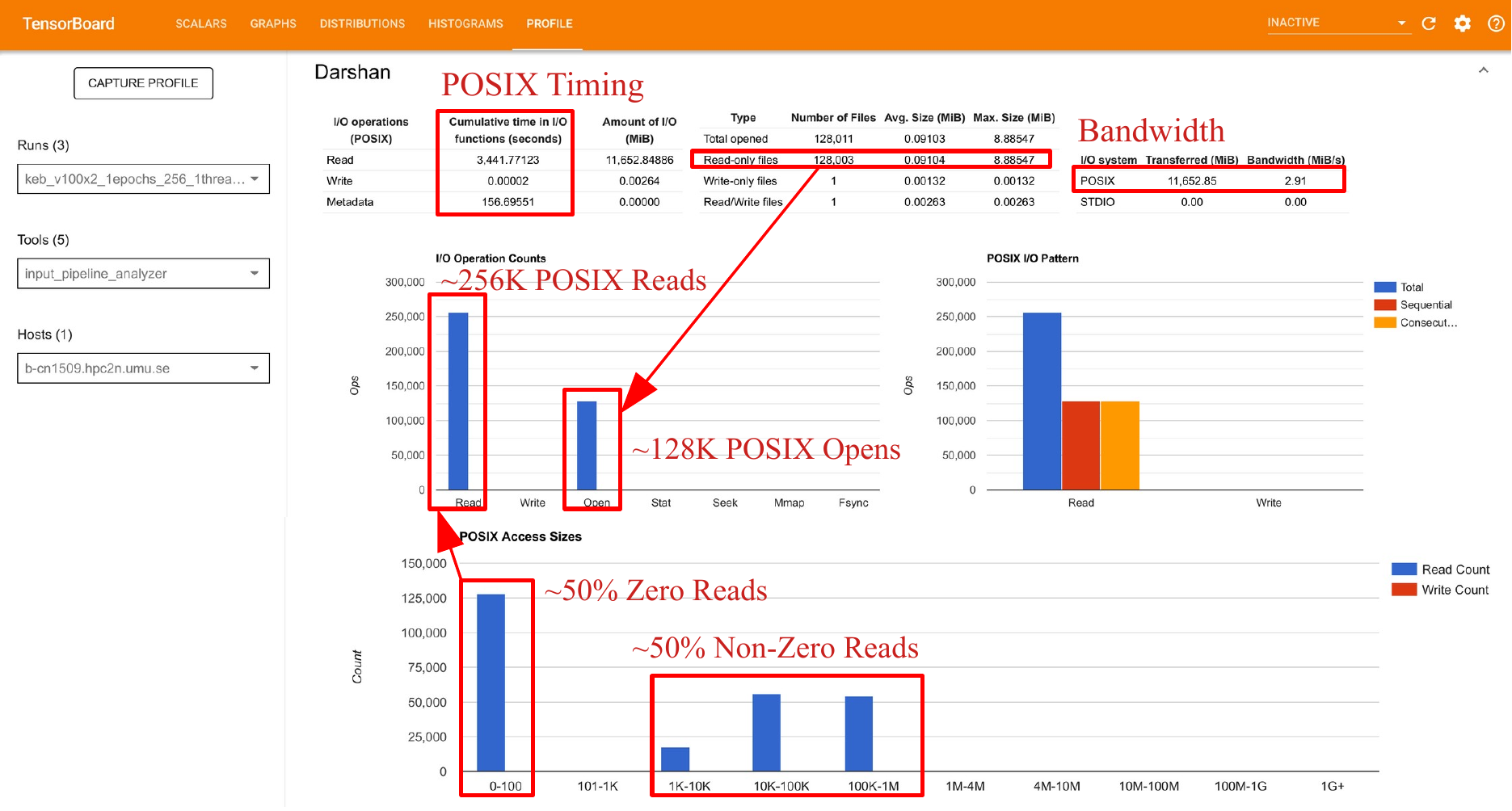}
		\caption{An extract tf-Darshan visualization for ImageNet training, showing activities on POSIX I/O operation counters, access pattern, and access sizes. The profiling results show twice the number of read operations relative to files opened. Approximately 50\% of the reads are small reads. Upon examining the TraceViewer, it can be seen all POSIX read operations are followed by a zero-length read.}
		\label{fig:keb-imagenet-tensorboard-unoptimized}
	\end{subfigure}
	\begin{subfigure}{\linewidth}
		\centering
		\includegraphics[width=\linewidth]{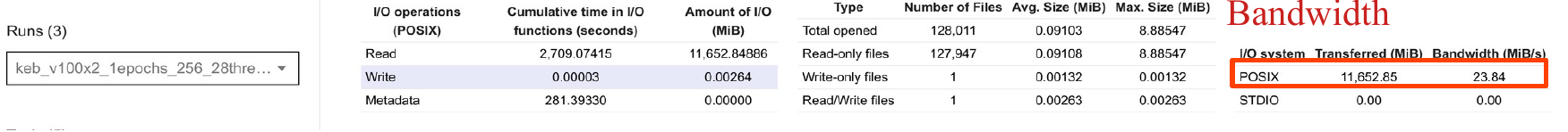}
		\caption{ImageNet training with increased threading to execute input-pipeline (from one to 28 threads) results in higher bandwidth, from approximately 3~MB/s to 24~MB/s. An increase in approximately $8\times$.}
		\label{fig:keb-imagenet-tensorboard-optimized}
	\end{subfigure}
	\caption{tf-Darshan profiling results for Image classification use-case.}
\end{figure*}

\begin{figure}[t]
	\centering
	\includegraphics[width=\linewidth]{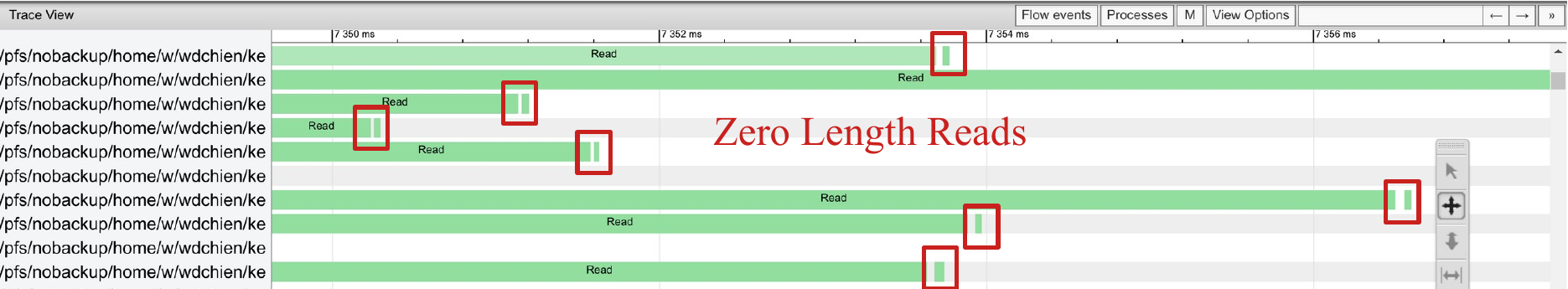}
	\caption{An extract of a number timelines of TraceViewer for ImageNet training, showing activities POSIX I/O. Each file is represented as one timeline. Tracing shows that each file read ends with a zero-length read, thus explaining the 50\% extra read operation relative to the number of files opened.}
	\label{fig:keb-imagenet-traceviewer}
\end{figure}

Image classification is a representative ML workload. We classify images in ImageNet with a batch size of 256 and 500 steps. The total size is approximately 11.6~GB (128K files) with a median size of 88~KB. We specify one thread to execute the preprocessing pipeline, together with a prefetch buffer of 10 batches. We execute this test on one Kebnekaise node using two NVIDIA V100 GPUs. To examine training performance, we ask TensorFlow to profile a full epoch and visualize the results on TensorBoard in Fig.~\ref{fig:keb-imagenet-tensorboard-unoptimized}.

According to TensorFlow Profiler, the training is highly input bounded. Approximately 96\% of the sampled step time is to wait for input data. tf-Darshan reports that the training has a very low POSIX bandwidth at approximately 3~MB/s, indicating a very heavy bottleneck. It reports that approximately 128K files were opened on the POSIX layer, and over 256K POSIX reads were performed. This is approximately doubled the number of files opened. At the same time, 50\% of the reads are neither sequential nor consecutive. By examining the POSIX read size, 50\% of the read calls have a length below 100~Bytes. The rest of the reads are of length 1~KB to 1~MB. These small-sized reads attribute to the low I/O performance.

We follow up on these observations by using the TraceViewer tool in TensorFlow Profiler, which visualizes tracing of TensorFlow operations and POSIX operations extracted by tf-Darshan. In the tf-Darshan panel, each line represents a file recorded by tf-Darshan and shows its respective operations. By zooming into the POSIX read activity reported by tf-Darshan in Fig.~\ref{fig:keb-imagenet-traceviewer} (an extract of relevant timelines), we notice that most read operations are one-off, meaning that a single I/O operation consumes the entire file. At the same time, they are all followed by a POSIX read with length zero. These measurements would explain our observation of POSIX operation counts and read size distribution. Upon examining the TensorFlow source code, the read file operation consists of a loop that performs \texttt{pread}. The function returns only upon \texttt{pread} returning zero, which signals the end of the file.

We try to optimize and rerun the use-case with threading to use 28 threads in \textsf{\small{tf.data.Dataset.map}} for parallel data preprocessing and present the result in Fig.~\ref{fig:keb-imagenet-tensorboard-optimized}. By increasing the number of threads, it is possible to increase the bandwidth to close to 24~MB/s. While still a low performance, it is an increase of approximately $8\times$ relative when only one thread is used.

\subsection{Malware Detection} 
\begin{figure*}
	\centering
	\includegraphics[width=\linewidth]{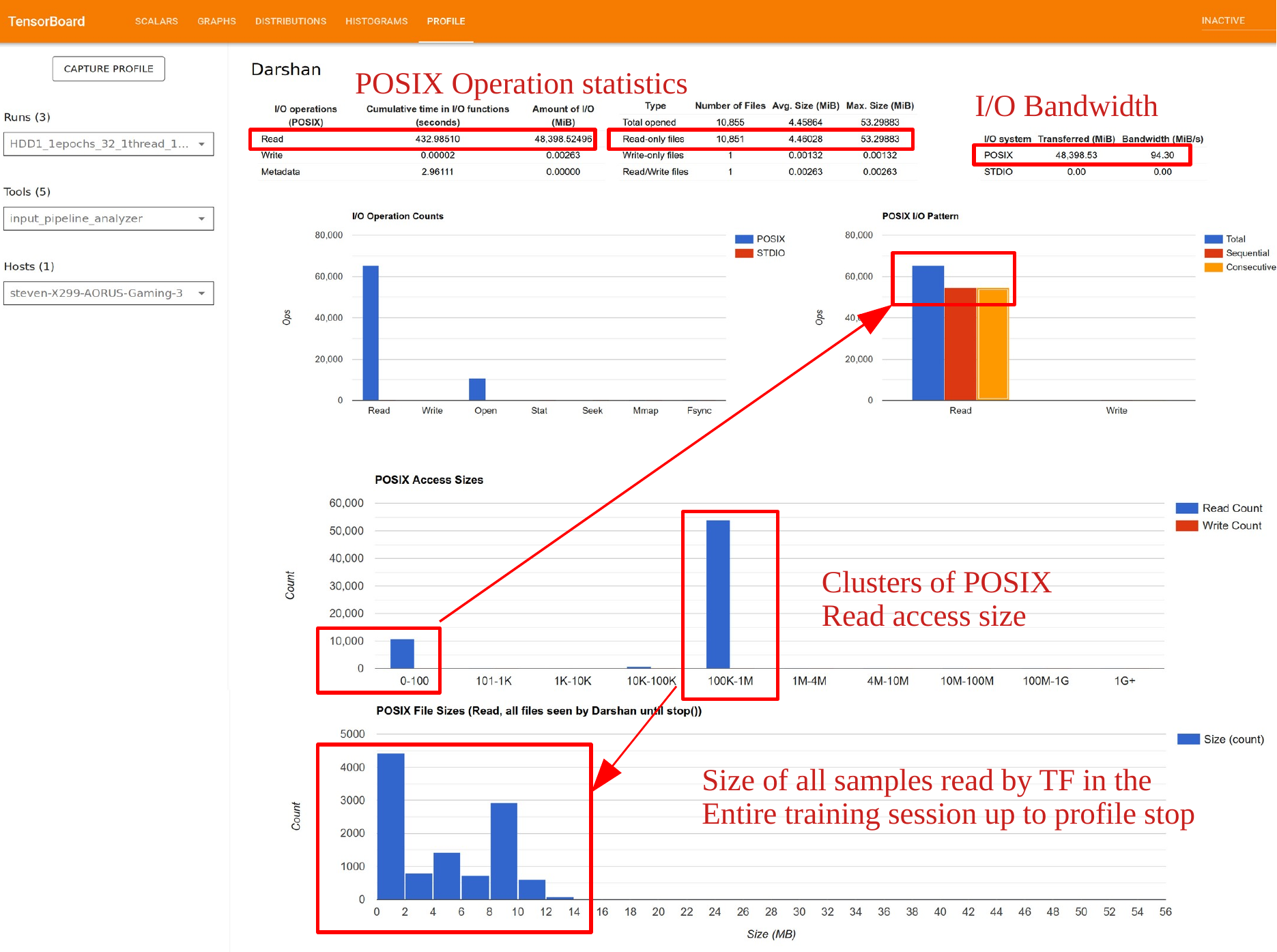}
	\caption{Result of profiling training of one epoch in the Malware Classification Challenge visualized through tf-Darshan's TensorBoard extension. Unlike ImageNet, the number of zero-length reads are relatively small, with a majority of reads clusters between 100KB to 1MB and a larger number of sequential consecutive reads. At the same time, the bandwidth is significantly higher.}
	\label{fig:malware-tensorboard-unoptimized}
\end{figure*}

\begin{figure}[t]
	\centering
	\includegraphics[width=\linewidth]{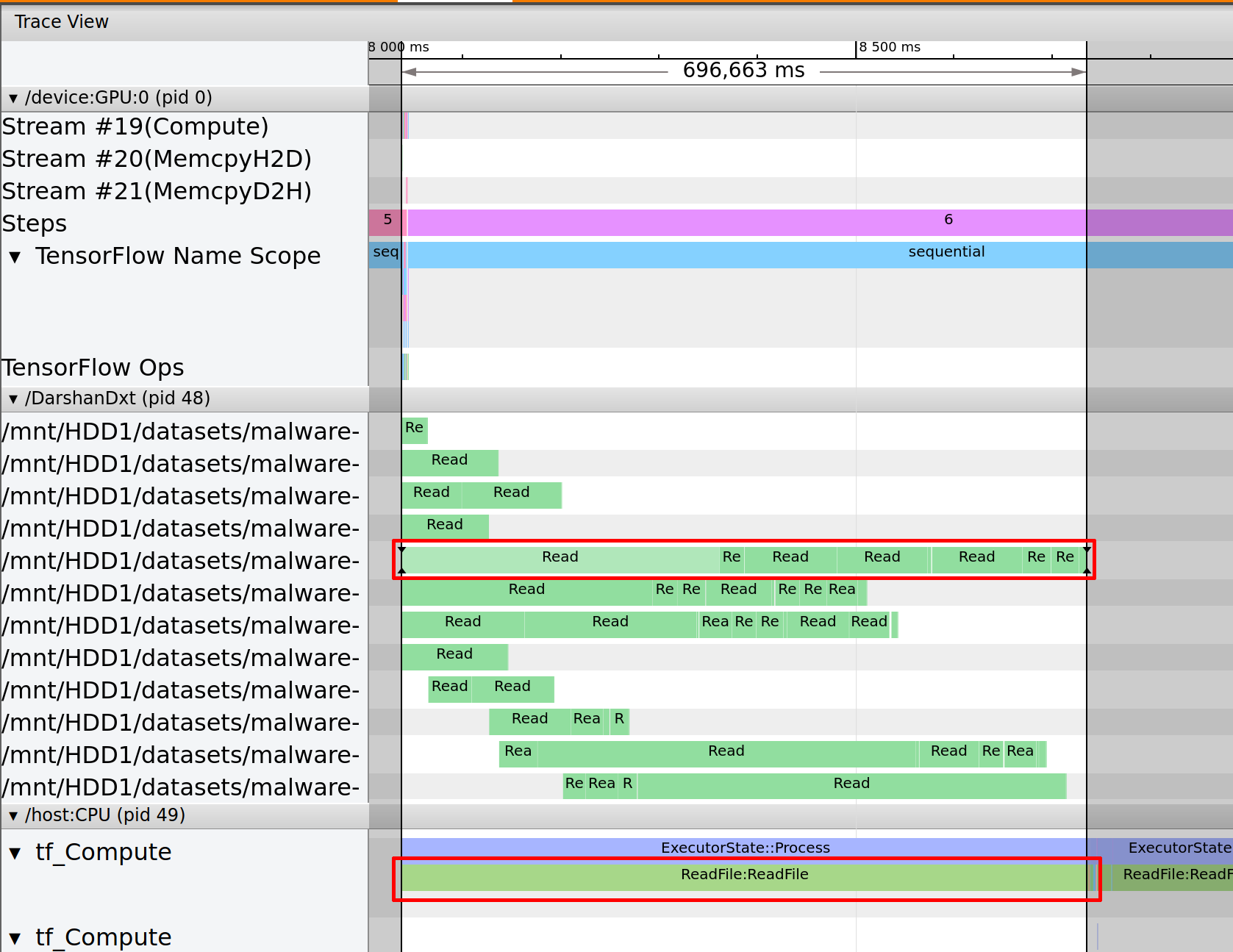}
	\caption{An extract of relevant timelines in TraceViewer for Malware Detection training profiling. The POSIX read operations that are triggered by TensorFlow's \emph{ReadFile} operation can be inferred by their respective range, as annotated in red.}
	\label{fig:malware-traceviewer}
\end{figure}

Our Malware classification use-case classifies executable byte codes through a CNN by opening and formatting bytecode data as grayscale images. The dataset consists of 10,868 bytecode files with a median size of approximately 4~MB and approximately 48~GB in total. Compared to our ImageNet Classification use-case, the dataset in this use-case has fewer samples, while individual samples are significantly larger. A relatively shallow network and large file size can imply a stressful workload for the I/O system. We perform training of one epoch with a batch size of 32 and specifying one thread to execute the preprocessing pipeline. We also ask for a prefetch buffer of 10 batches. Similar to the ImageNet use-case, we profile the full epoch.

As the use-case features larger training samples and a CNN that is much smaller than AlexNet, it is reasonable to expect that I/O operations will dominate execution. The TensorFlow Profiler's step time breakdown analysis, where 99\% of the sampled time is for waiting for input data, verifies this expectation. Unlike the ImageNet use-case, the GPU device computes time is negligible, meaning that the training is purely I/O-bound. According to tf-Darshan's I/O operation count statistics in Fig.~\ref{fig:malware-tensorboard-unoptimized}, over 60,000 POSIX read operations are performed and bandwidth of 94~MB/s is reported. By inspecting POSIX read size distribution, it is clear that the majority of reading sizes are within one MB. Given the median size of the dataset is four MB, it implies that most of the read operations occur in segments. This result is confirmed by the POSIX I/O pattern where the majority of reading operations are both sequential and consecutive. Since we only perform sequential read on data samples and given our observation of zero-sized read in the ImageNet use-case, the rest of the read operations are likely reads that return zero sizes. By zooming into POSIX operation traces in the TraceViewer, we can confirm this result.

\begin{figure*}
    \begin{subfigure}{\linewidth}
        \centering
        \includegraphics[width=\linewidth]{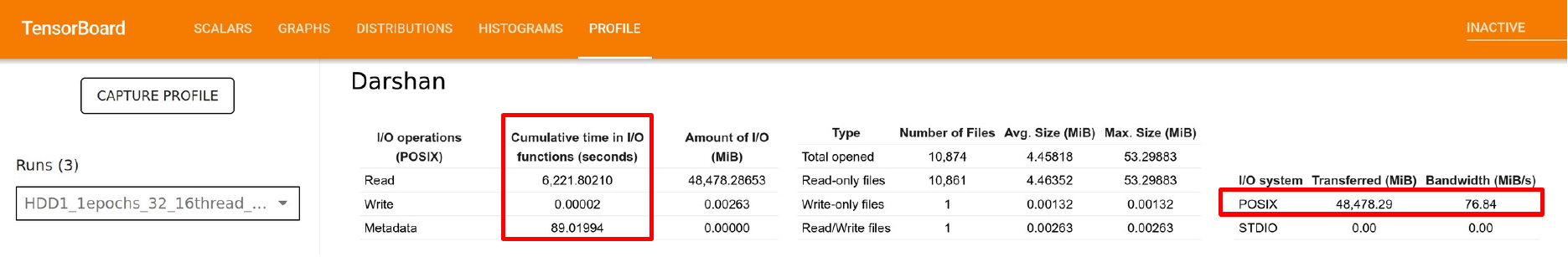}
        \caption{Change in POSIX bandwidth after increasing the number of threads which results in worse performance, with a drop from approximately 94~MB/s to 77~MB/s.}
        \label{fig:malware-tensorboard-threads-not-helping}
    \end{subfigure}
    
    \begin{subfigure}{\linewidth}
        \centering
        \includegraphics[width=\linewidth]{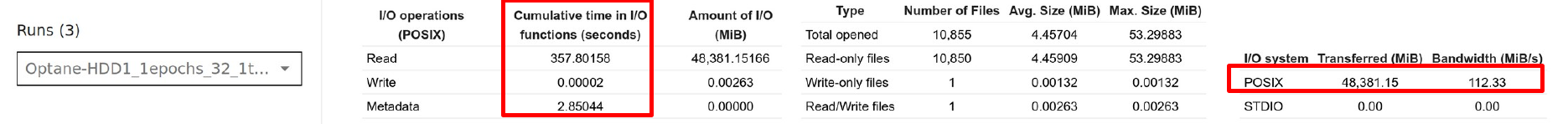}
        \caption{Change in POSIX bandwidth after manually moving all the files that are smaller than 2~MB to Intel Optane SSD. While the staged data only accounts for 8\% of the dataset size, it results in 19\% improvement in bandwidth.}
        \label{fig:malware-tensorboard-optane-helping}
    \end{subfigure}
    \caption{Different parameters for Malware Classification training.}
\end{figure*}
\begin{figure*}[t]
    \centering
    \includegraphics[width=\linewidth]{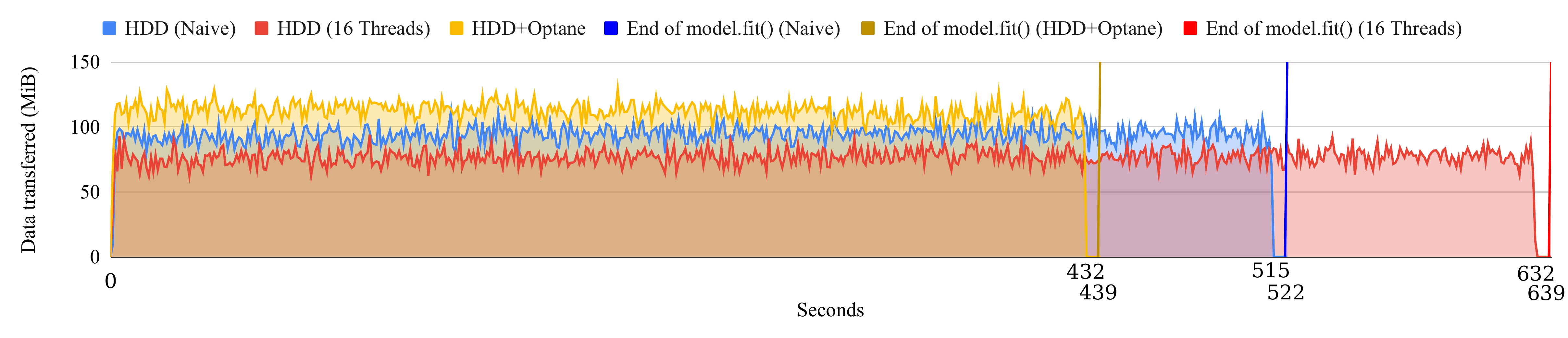}
    \caption{Disk activities collected by Dstat in the background while the un-optimized, threaded, and optimized parameters are being used for Malware Classification training.}
    \label{fig:malware-dstat}
\end{figure*}

To examine the bottleneck of the training, we rerun the case and profile for a small number of steps to inspect the tracing through the TraceViewer. We extract the relevant timelines and show them in Fig.~\ref{fig:malware-traceviewer}. By comparing the time range using the TraceViewer, it is possible to infer the relevant POSIX file operations executed by a TensorFlow \emph{ReadFile} operation. We examine the traces, find that the files are read in a number of the segment, and the reading time varies by a magnitude of milliseconds. However, the metadata of the traces shows that they are reading the same length, implying a large variance between the time taken to read each segment. At the same time, the time taken to execute TensorFlow operations on the GPU takes almost negligible time, suggesting that it is unlikely to be able to achieve complete overlapping of input preprocessing and training. We attempt to optimize the training by increasing the number of threads performing I/O operations in \textsf{\small{tf.data.Dataset.map()}} to 16 and use hyper-threading. However, the increase of threads results in a decreased bandwidth as shown in Fig.~\ref{fig:malware-tensorboard-threads-not-helping}. One explanation is the relatively large file sizes in the use-case, with a median size of approximately 4~MB. An increase in the numbers of parallel read can result in a higher bandwidth contention. Our image classification use-case, on the other hand, has small file sizes, with media size approximately 88~KB.

By investigating the file size and POSIX read size distribution, it is clear that files below 1~MB can be read completely in one single read operation. With approximately 4,420 files below 2~MB, they can account for $4420\times2MB=8.6GB$ of data in a worst-case scenario, which accounts for only approximately  18\% of the total size. We inspect the sizes of those files and find that they account for only 3.7~GB, which is approximately 8\% of the total size; at the same time, it accounts for 40\% of the total number of files. Knowing that HDD is good in sequential read rather than random read, we move all those files into our Intel Optane SSD, which provides high-speed data access. We rerun the training and profile for one epoch using one thread. The result in Fig.~\ref{fig:malware-tensorboard-optane-helping} shows that by staging small files, using the information we obtain from tf-Darshan, we can increase the POSIX bandwidth by approximately 19\%. While it is possible to obtain purely file size information without using tf-Darshan, gaining additional knowledge on POSIX reading size and other trace information provides a bigger picture and helps us derive a decision that minimizes storage space requirement on a fast storage tier. For example, in a case where there is a small number of larger files, one might intuitively stage the larger files to faster access, which in the end may not provide a big improvement to performance as a large number of smaller reads remain. Finally, we quantify the disk activities when using the three configurations using Dstat in Fig.~\ref{fig:malware-dstat} and show that our optimized version uses the highest bandwidth and lowest training time.

%% file: rel-work.tex
\section{Related Work}
\label{sec:rel-work}
Deep learning applications are emerging as a critical class of applications on HPC systems, either as standalone applications such as drug discovery, antibody designs, anomaly detection, or as a part of complex systems. Extensive works~\cite{jouppi2017datacenter, gu2017deepprof, kwon2019understanding} have studied the computation requirements, data reuse, and memory access patterns in deep learning applications. Recent works~\cite{chien2018characterizing, choi2019faster} have shown that the increased computing capability may reduce the chance of overlapping I/O operations because the computation time is reduced. Several works~\cite{chowdhury2019characterization} have studied the I/O patterns in machine learning workloads on specific parallel file systems such as BeeGFS and Lustre. These works focus on performance characterization and used customized analysis approaches. Our work focuses on an I/O profiling tool to pinpoint bottlenecks in the most popular deep learning framework for performance optimization.
To our knowledge, it is the first work that provides profiling and trace capability and visualization in TensorBoard.  

Chowdhury et al.~\cite{chowdhury2019characterization} characterized the performance of a new parallel file system called BeeFS. They studied the I/O performance in two deep learning frameworks, i.e., the ImageNet data reader in TensorFlow and the classification in LBANN~\cite{van2015lbann} framework. Their work used the default Darshan profiler for large parallel file systems and performance analysis of the logs. 

Many works~\cite{yu2008performance, patel2019revisiting, patel2020uncovering} have characterized the performance of various file systems and their impact on general applications. Yu et al.~\cite{yu2008performance} characterized Lustre parallel file system atop a Cray XT supercomputer. Patel et al.~\cite{patel2019revisiting} performed an in-depth statistical analysis of a year-long collection of storage data to study the patterns on accessing large parallel file systems on HPC systems. Patel et al.~\cite{patel2020uncovering} classifies files and then studied the file reuse and access characteristics in files on petabyte-scale systems on production supercomputers. Our profiling tool takes a different perspective from these file system based characterizations, but focus on the I/O requirement in the deep learning workloads. 

There are several profiling tools for characterizing computation and memory access in deep learning applications. For instance, DeepProf~\cite{gu2017deepprof} provides a GPU trace analysis tool to characterize the performance of TensorFlow applications. The TensorFlow framework has built-in profiling capabilities~\cite{tfprofiler}, such as collecting hardware counters and also supports plugin extensions. Our tool provides I/O profiling capability that is currently unavailable in other toolsets. It leverages the TensorBoard~\cite{tfboard} visualization toolkit in TensorFlow to provide a consistent interface to users.

%% file: conclusion.tex
\section{Discussion and Conclusion}
\label{sec:conclusion}
In this paper, we developed tf-Darshan, a TensorFlow profiler and tracer that can capture and analyze fine-grained I/O information in ML workloads by leveraging the capability of Darshan. We illustrated using a STREAM-like benchmark, the possibility of obtaining profiling data while TensorFlow is executing. The tf-Darshan system adds a moderate (10\%-20\%) overhead, where the data post-processing (and not the execution itself) contributes most to the overhead cost. The results provided by Darshan and visualized through TensorBoard guided the optimization of our use-cases, either by increasing parallelism or by staging a subset of data into a faster storage tier. Our malware classification training resulted in a 19\% improvement of bandwidth, by staging data samples that accounts for 8\% of total data size to the Intel Optane SSD.

Apart from exposing Darshan's capability for ML I/O optimization, our work shows the opportunity of using Darshan as a profiling library for applications, similar to the CUDA CPUTI for NVIDIA GPUs. Once introducing the capability of runtime attachment, Darshan has the capability of providing information for such as auto-tuning during execution. For instance, one observation from our profiling on TensorFlow is a large number of small reads, which is consistent with previous works~\cite{chowdhury2019characterization}. By understanding information on file sizes in relation to POSIX read sizes distribution, it is possible to reduce small reads. TensorFlow already uses auto-tuning extensively in many different aspects, from setting up computation kernels to determining parallelism. The information from tf-Darshan has the potential of improving this process with I/O specific information. Even though our work focuses on a non-trivial and loosely coupled integration of Darshan with TensorFlow, the same technique can potentially be applied to other ML frameworks. PyTorch~\cite{paszke2019pytorch} for example, also provides a performance profiler~\cite{pytorch-profiler} in the Automatic Differentiation Package~\cite{paszke2017automatic}. Despite the specific implementations of I/O operations in different ML frameworks, they share the general data ingestion pattern of independent I/O in parallel. One notable exception is \emph{Caffe}, which uses LMDB, a memory-mapped database through \texttt{mmap}. Currently, Darshan's POSIX module can capture \texttt{mmap} operations but requires extensions to further capture fine-grained interactions, e.g., \texttt{msync} calls.

I/O has long been an optimization target in HPC. ROMIO, a major achievement in implementing MPI-IO for example, uses \emph{data sieving} and \emph{two-phase I/O} to reduce the number of metadata operations and to increase the I/O size when performing non-contiguous access~\cite{thakur1999data}. ML workloads on the other hand only perform individual I/O of data samples and parallel I/O in the ML context commonly refers to executing multiple data producers in parallel. While effective in some cases (where our image classification achieved $8\times$ increase in bandwidth after applying threading), small file sizes can severely impact the performance (as seen in the image classification case-study), especially in parallel file systems where metadata operations are expensive. One way to improve bandwidth performance is to use data containers such as TFRecord~\cite{tfrecord} that contains multiple data samples. However, the preparation of such containers still requires a separate preprocessing step with I/O for each sample.

While emerging storage technologies~\cite{brinkmann2020ad,zhu2019efficient,wei2018exploring} that aim to elevate I/O bottlenecks, and projects such as SAGE~\cite{narasimhamurthy2018sage} and DAOS~\cite{breitenfeld2017daos}, are being intensively studied, I/O will likely remain a challenging aspect when deploying ML workload. By enabling fine-grained profiling and tracing capability, we also enable the opportunity for automated decision making and auto-tuning in the future.

In the future, the profiler can be optimized to reduce the overhead; for instance, detailed timeline tracing can be optionally discarded if not required. We currently require only three extra functionalities in Darshan, namely, Extraction of: Darshan runtime structure that provides a file counts; and Darshan module buffer, where the counters of individual files are stored. As for our TensorFlow modules, a future work is to package it as a TensorFlow module plugin through refactoring, in order to contribute the tool to a wider audience.